\documentclass[prl,aps,twocolumn, a4paper,longbibliography]{revtex4-2}

\usepackage[paperwidth=210mm,paperheight=297mm,centering,top=2cm,bottom=2.8cm,right=1.8cm,left=1.8cm]{geometry}

\pdfoutput=1
\date{\today}

\usepackage{xspace}
\usepackage{soul}

\newcommand{\um}{\upmu{\rm m}}
\newcommand{\nK}{\textrm{nK}}
\newcommand{\kB}{k_{\textrm{B}}}

\newcommand{\Tc}{T_{\textrm {c}}}

\newcommand{\upd}{\textrm{d}}
\newcommand{\potassium}{^{39}\textrm{K}}

\newcommand{\kxi}{k_\xi}

\newcommand{\kp}{k_{\textrm{p}}}

\usepackage{float}
\usepackage{fourier}

\usepackage[dvipsnames]{xcolor}
\usepackage{graphicx}
\usepackage{amsmath,amssymb}

\usepackage{times}
\usepackage{upgreek}
\usepackage{psfrag} 
\usepackage{latexsym} 
\usepackage{amstext}
\usepackage{amsxtra} 

\usepackage{lineno}

\usepackage{textcomp}
\usepackage{amsfonts}
\usepackage{graphicx}
\usepackage{bm}
\usepackage{color}
\usepackage{braket}
\usepackage{siunitx}
\usepackage[svgnames]{xcolor}
\usepackage[normalem]{ulem}

\definecolor{myColor}{rgb}{0.02,0.12,0.3}
\definecolor{myciteColor}{rgb}{0.39,0.7,0.89}
\definecolor{myBlue}{rgb}{0.1,0.1,0.8}
\usepackage[colorlinks=true,citecolor=myColor,linkcolor=myColor,urlcolor=myColor]{hyperref}

\def\be{\begin{equation}}
\def\ee{\end{equation}}

\makeatletter
\def\@fnsymbol#1{\ensuremath{\ifcase#1\or *\or \dagger\or \ddagger\or
   \mathsection\or \mathparagraph\or \|\or **\or \dagger\dagger
   \or \ddagger\ddagger \else\@ctrerr\fi}}
\makeatother

\usepackage{titlesec}

\begin{document} 

\title{
Weak wave turbulence as a precursor to universal coarsening in a homogeneous Bose gas\\
}
\author{Simon~M.~Fischer$^{\ast}$}
\author{Martin~Gazo}
\author{Sebastian~J.~Morris}
\author{Nikolai~Maslov}
\author{Haoyu~Zhang}
\author{Ji\v r\'i Etrych}
\author{Gevorg~Martirosyan}
\author{Christoph~Eigen}
\author{Zoran~Hadzibabic}
\affiliation{Cavendish Laboratory, University of Cambridge, J. J. Thomson Avenue, Cambridge CB3 0US, United Kingdom}

\begin{abstract}
Relaxation and condensation of an isolated low-energy Bose gas provide an ideal setting for the study of the universal features of far-from-equilibrium many-body dynamics and the emergence of long-range order. Conceptually, the emergence of such order involves two steps: the formation of local coherence, on a system-specific microscopic lengthscale, and the spreading of coherence, over lengthscales much larger than any microscopic scale. The latter is understood in terms of universal phase-ordering kinetics, or coarsening, characterized by an algebraic growth of the coherence length. Here, for a homogeneous Bose gas with tunable interactions, we show that the former also has a universal description, within the framework of weak wave turbulence (WWT). Specifically, the initial transport of particles to low momenta corresponds to an inverse turbulent cascade that is, in agreement with the WWT theory, characterized by a power-law momentum distribution, with exponent $\gamma = 2.4(1)$, and transport times ${\propto} (na)^{-2}$, where $n$ is the gas density and $a$ the $s$-wave scattering length. 
\end{abstract}

\maketitle

A Bose--Einstein condensate is a paradigmatic state of matter with long-range coherence, and the formation of such order is relevant across many fields and lengthscales~\cite{Kibble:1976,Zurek:1985,Berges:2008,Moore:2016,Proukakis:2017,Berges:2021}.
Theoretically, this process has been linked to various forms of wave and vortex turbulence~\cite{Kraichnan:1967,Svistunov:1991,Dyachenko:1992,Kagan:1994,Semikoz:1995,Kagan:1995ch,Svistunov:1995,Semikoz:1997,Lacaze:2001,Berloff:2002,Connaughton:2004,Nazarenko:2011,Semisalov:2021,Zhu:2022,Zhu:2023b,Barenghi:2023b,Rosenhaus:2025,Zakharov:2025}, which are also at the heart of recent theories of nonthermal fixed points and universality far from equilibrium~\cite{Berges:2008,Nowak:2012,Chantesana:2019,Mikheev:2023}.
Experimentally, there is a long history of studies of condensation dynamics in ultracold atomic gases, both in harmonic traps~\cite{Miesner:1998b,Kohl:2002,Ritter:2007,Hugbart:2007,Smith:2012,Moreno-Armijos:2025} and in optical box traps~\cite{Glidden:2021,Gazo:2025,Martirosyan:2025,Morris:2026}, the latter allowing closer connections with theory and other physical systems.

\begin{figure}[b!]
\centerline{\includegraphics[width=\columnwidth]{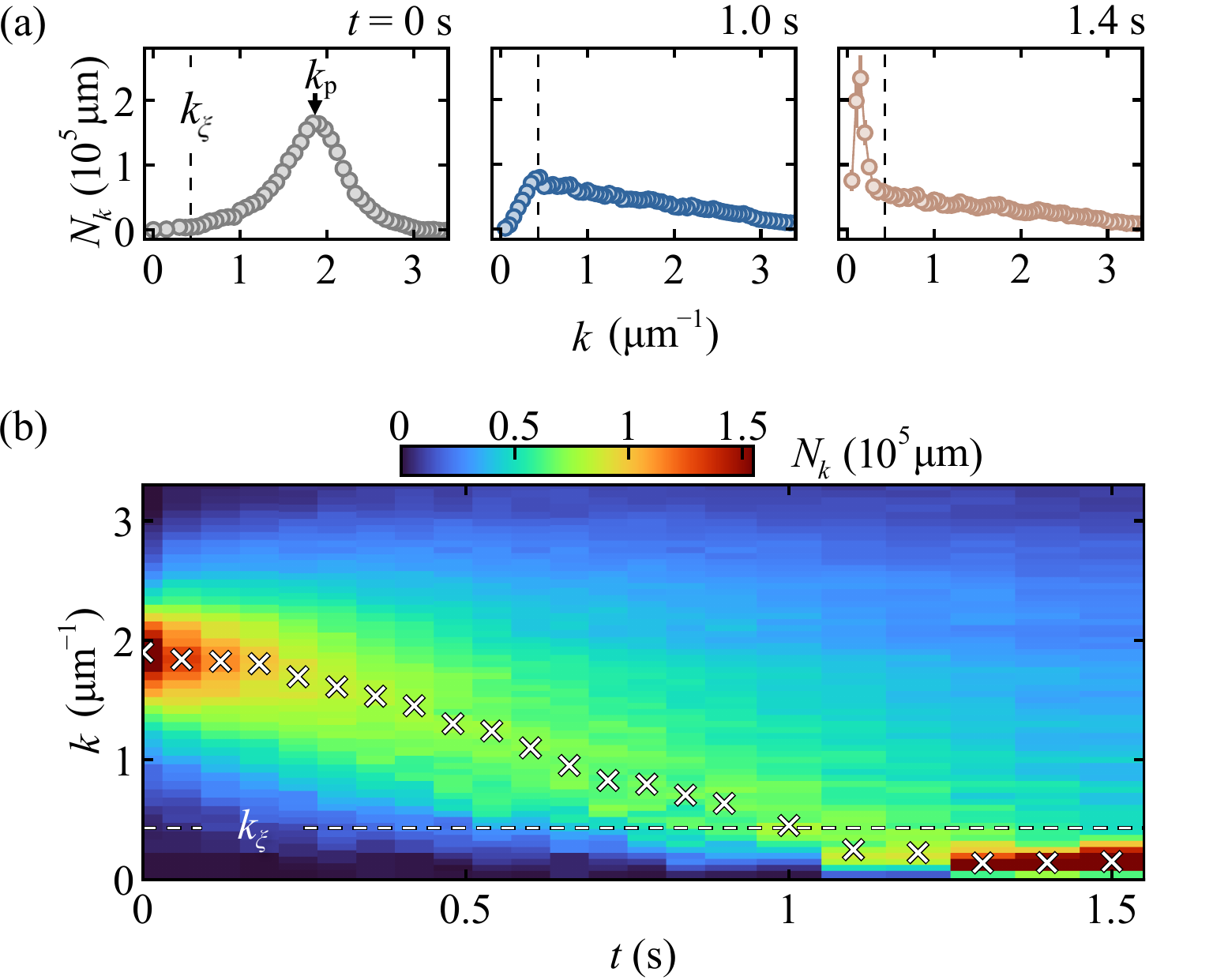}}
\caption{
Momentum-space particle transport during far-from-equilibrium condensation in an isolated homogeneous Bose gas. The system starts (at time $t=0$) in an incoherent low-energy state and condenses as it relaxes towards equilibrium. We show the evolution of the spectral population density $N_k(k) = 4\pi k^2 n_k(k)$, where $n_k$ is the momentum distribution; $\kp$ is the peak position of $N_k$, and $\kxi$ the inverse healing length. Here the gas density is $n\approx2.8\,\um^{-3}$ and the scattering length $a =50\,a_0$, corresponding to $\kxi \approx 0.43\,\um^{-1}$.
(a) The condensate (spectral peak near $k=0$) starts to form only after $\kp$ drops to $\approx \kxi$.
Here, we study the initial particle transport, to $\approx\kxi$, and show that it corresponds to weak wave turbulence (WWT).
(b) Summary of the evolution of $N_k$; white crosses indicate $\kp(t)$.}
\label{fig:1}
\end{figure}

A homogeneous Bose gas that is initially far from equilibrium and incoherent, but has sufficiently low energy, can condense in isolation. The emergence of real-space coherence during the relaxation towards equilibrium is simply related (by a Fourier transform of the momentum distribution $n_k$) to the transport of the majority of particles towards low momenta~\cite{Kraichnan:1967,Svistunov:1991,Dyachenko:1992,Chantesana:2019,Glidden:2021}. The interactions that drive this relaxation introduce a characteristic microscopic lengthscale, the healing length $\xi = 1/\sqrt{8\pi n a}$, where $n$ is the gas density and $a$ the $s$-wave scattering length, and one can conceptually separate the relaxation dynamics into two regimes.
We illustrate this (with our experimental data) in Fig.~\ref{fig:1}, where we show the evolution of the spectral population density, $N_k(k) = 4\pi k^2 n_k(k)$, which has a well-defined peak position $\kp$.
First, the particles undergo transport to relatively low momenta, 
such that $\kp \approx \kxi = 1/\xi$, and then the condensate starts to grow~\cite{Kagan:1994,Kagan:1995ch,Berloff:2002,Proukakis:2017,notequasicondensate}.

Recent homogeneous-gas experiments~\cite{Martirosyan:2025,Morris:2026} have focused on the late-time emergence of long-range order. They have shown that once the coherence length $\ell$ significantly exceeds $\xi$, the gas exhibits universal coarsening that is independent of the strength of the interparticle interactions~\cite{Martirosyan:2025}, and that microscopically this process is linked to the decay of a tangle of quantized vortex filaments~\cite{Morris:2026}, known as Vinen turbulence~\cite{Svistunov:1995,Berloff:2002,Vinen:1957,Vinen:1957c}.

\begin{figure*}[t!]
\centerline{\includegraphics[width=\textwidth]{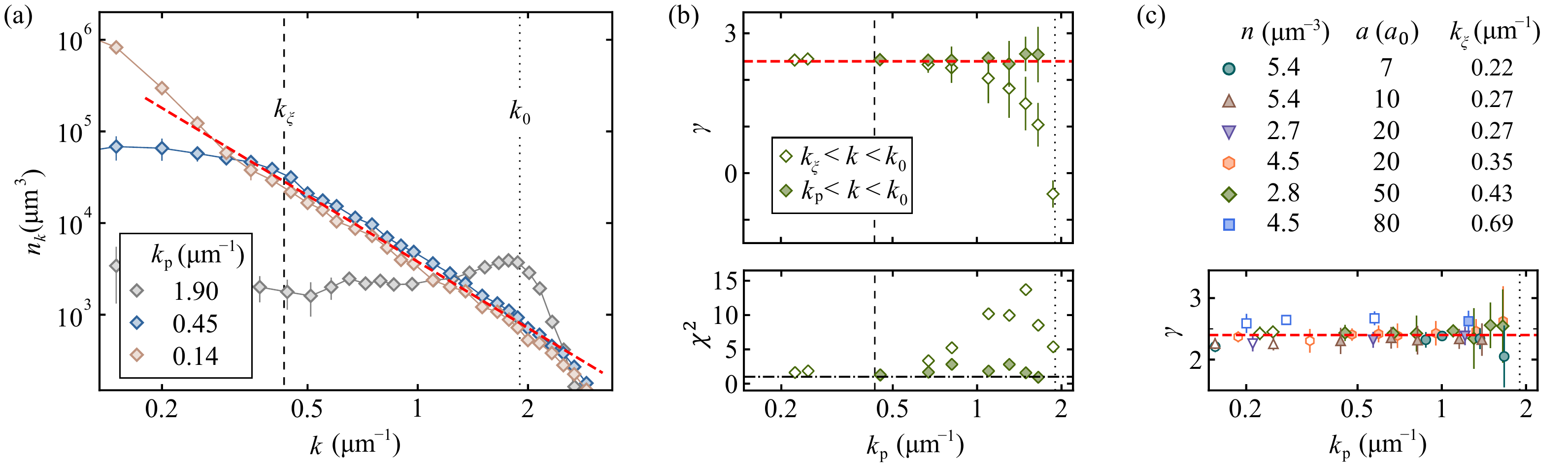}}
\caption{
Emergence of the WWT power-law spectrum. 
(a) Momentum distribution $n_k$ for various relaxation times, corresponding to different $\kp$, for $n\approx2.8\,\um^{-3}$ and $a =50\,a_0$, so $\kxi \approx 0.43\,\um^{-1}$;  $k_0 \approx 1.9\,\um^{-1}$ is the initial value of $\kp$. Once $\kp$ drops to $\approx \kxi$, in the spectral range $\kxi \lesssim k \lesssim k_0$ we observe a power-law $n_k \propto k^{-\gamma}$, with $\gamma \approx 2.4$ (red dashed line), close to theoretical predictions (see text).
(b)~Fitted $\gamma$ values and the corresponding $\chi^2$ (with the dot-dashed line showing $\chi^2 = 1$) for fits in two different spectral ranges: from $\kxi$ to $k_0$ (open symbols) and from the instantaneous $\kp$ to $k_0$ (solid symbols). In the former case, the fitted $\gamma$ approaches $2.4$ (horizontal dashed line) as $\kp$ approaches $\kxi$ (vertical dashed line), and then remains $\approx 2.4$ as $\kp$ decreases further. In the latter case, we always observe $\gamma\approx 2.4$, which shows that the power-law spectrum forms in the wake of the decreasing $\kp$.
(c) 
Different $n$ and $a$. Fitting from $\kp$ to $k_0$ for $\kp > \kxi$ (solid symbols) and from $\kxi$ to $k_0$ for $\kp < \kxi$ (open symbols), we always observe $\gamma \approx 2.4$ (dashed line).
}
\label{fig:2}
\end{figure*}

In this Letter, we study the momentum-space particle transport that precedes the emergence of long-range order, and show that it has a universal description, in terms of weak wave turbulence (WWT).
Starting from a low-energy far-from-equilibrium state, $n_k$ evolves into a power-law, characteristic of a turbulent cascade. Specifically, we observe a cascade exponent $\gamma = 2.4(1)$, which is consistent with theoretical predictions for WWT in a homogeneous gas with high mode occupations; the analytical (Kolmogorov--Zakharov) prediction is $\gamma = 7/3$~\cite{Svistunov:1991,Dyachenko:1992,Nazarenko:2011,Zakharov:2025}, while numerical results vary in the range $2.44\textup{--}2.52$~\cite{Semikoz:1995, Semikoz:1997, Lacaze:2001, Connaughton:2004,Semisalov:2021,Zhu:2022,Zhu:2023b}. This power-law $n_k$ is observed for $k\gtrsim \kxi$ and emerges in the wake of the decreasing $\kp$.
We also show that in this regime the characteristic time for particle transport scales as $1/(na)^2$, which is distinct from the classical (collisional) Boltzmann scaling $1/(na^2)$ and agrees with~the~WWT~\mbox{theory}.

Our experiments are performed with $\potassium$ atoms in the lowest hyperfine state, held in a cylindrical box trap~\cite{Eigen:2016,Navon:2021} of volume $V \approx 5.5 \times 10^4 \,\um^{3}$ (radius $21(2)\,\um$ and length $40(4)\,\um$).
We vary the gas density in the range $n \approx (2.7 \textup{--} 5.4) \,\um^{-3}$, corresponding to the total atom number $N\approx(1.5 \textup{--} 3.0)\times 10^5$ and critical temperature for condensation  $\Tc \approx (80 \textup{--} 127) \, \nK$. We tune the $s$-wave scattering length
$a$ using the magnetic Feshbach resonance at $402.7\,\textrm{G}$~\cite{Etrych:2023}.
We prepare the gas in a far-from-equilibrium incoherent state~\cite{Martirosyan:2024,Martirosyan:2025} with energy per particle $\approx \kB \times 20\, \mathrm{nK}$, corresponding to equilibrium condensed fractions $\eta \approx 0.5 \textup{--} 0.6$.
The initial state is prepared at $a=0$ and we initiate relaxation by turning on interactions (setting $a>0$) at $t=0$.

In Fig.~\ref{fig:2}, we study the evolution of $n_k$ profiles for different $n$ and $a$. Our initial $\kp$ is equal to $k_0 \approx 1.9\,\um^{-1}$, and we choose $n$ and $a=(7\textup{--}80)\,a_0$ (where $a_0$ is the Bohr radius) such that $\kxi$ is always $\lesssim 0.7\,\um^{-1}$, 
allowing us to study dynamics in the spectral range $\kxi<k<k_0$.

In Fig.~\ref{fig:2}(a) we show, for $n\approx2.8\,\um^{-3}$ and $a =50\,a_0$, that once $\kp$ drops below $\kxi$, for $\kxi \lesssim k \lesssim k_0$ we observe a power-law $n_k \propto k^{-\gamma}$, with $\gamma\approx 2.4$ (red dashed line), close to the analytical WWT prediction $7/3$~\cite{Svistunov:1991,Dyachenko:1992,Nazarenko:2011,Zakharov:2025} and numerical results $2.44\textup{--}2.52$~\cite{Semikoz:1995, Semikoz:1997, Lacaze:2001, Connaughton:2004, Semisalov:2021,Zhu:2022,Zhu:2023b}.

In Fig.~\ref{fig:2}(b), we study how this power-law $n_k$ emerges. If we fit $\gamma$ for $\kxi \leq k \leq k_0$, we get values that gradually approach $\approx 2.4$ as $\kp$ approaches $\kxi$ (open symbols). On the other hand, if, for $\kp > \kxi$,  we fit $\gamma$ only from the instantaneous $\kp$ to $k_0$, we always observe $\gamma \approx 2.4$ (solid symbols), which shows that the spectrum with this $\gamma$ forms in the wake of the decreasing $\kp$. Note that for $\kp > \kxi$, the distribution is not really a power law in the full range $\kxi \leq k \leq k_0$, but our fits heuristically capture the approach to the WWT spectrum, whereas for $\kp \leq k \leq k_0$ the spectrum is always a power law; we show this by plotting (reduced) $\chi^2$ for both fitting ranges in the bottom panel of Fig.~\ref{fig:2}(b).

In Fig.~\ref{fig:2}(c), we summarize the results of such analysis for different $n$ and $a$. Here we show $\gamma$ values obtained by fitting from $\kp$ to $k_0$ as long as $\kp > \kxi$ (solid symbols) and from $\kxi$ to $k_0$ once $\kp < \kxi$ (open symbols). We always observe $\gamma \approx 2.4$ (dashed line) and get a combined estimate $\gamma = 2.4(1)$.

\begin{figure}[b!]
\centerline{\includegraphics[width=\columnwidth]{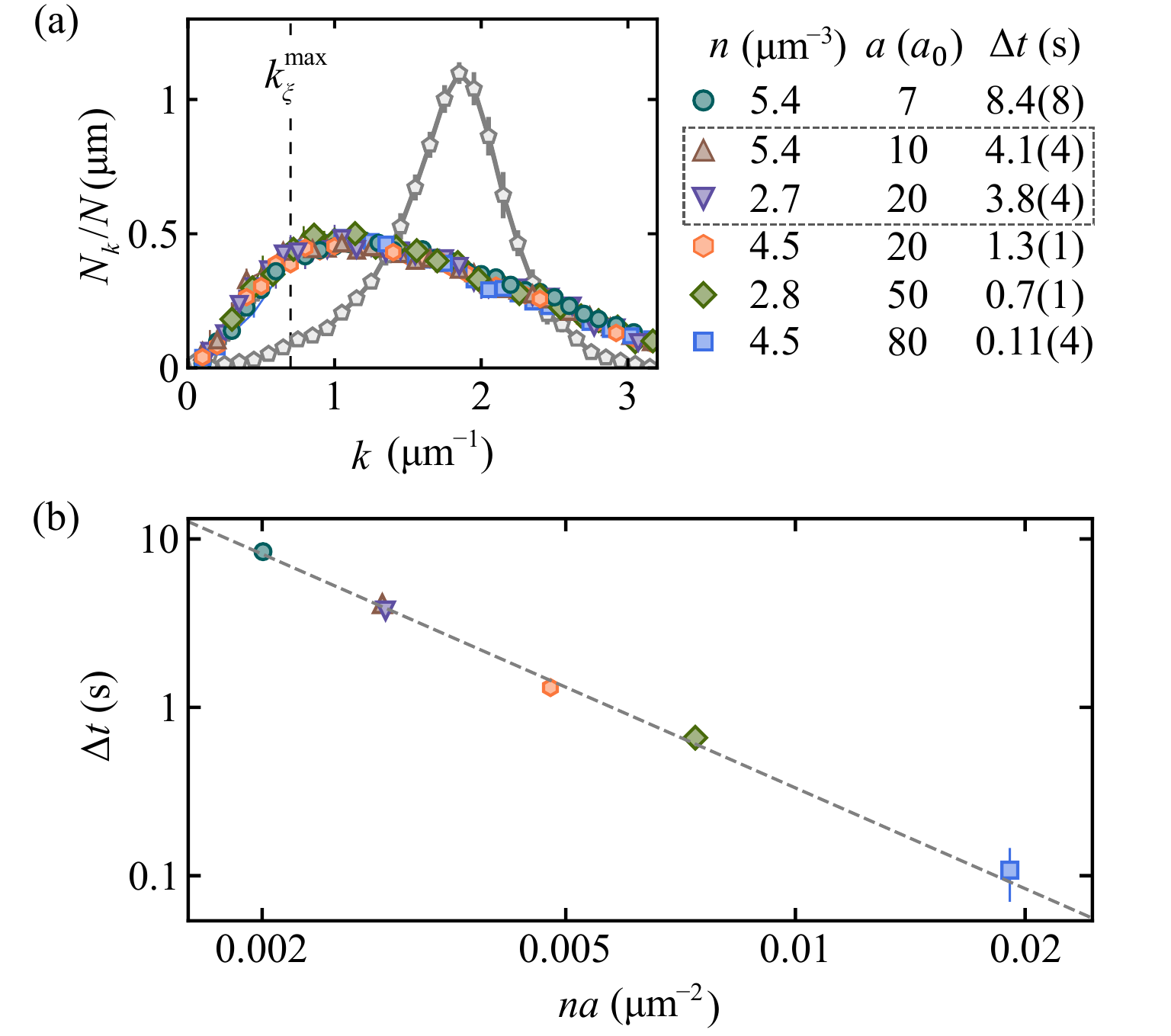}}
\caption{
Characteristic particle-transport time. 
For different $n$ and $a$ (as in Fig.~\ref{fig:2}), we extract the time $\Delta t (n,a)$ for $\kp$ to drop from $k_0 \approx 1.9\,\um^{-1}$ to $\approx k_0/2$.
(a) $N_k$ normalized by the total atom number $N$, for $\kp \approx k_0/2$ and different $(n,a)$ (colored symbols), and for our initial state (gray); the dashed line indicates the largest $\kxi$.
The fact that, as $\kp$ decreases, the $N_k/N$ curves for different $(n,a)$ remain the same shows that $\Delta t (n,a)$ fully captures the dependence of the dynamics on the interactions.
Importantly, $\Delta t$ is the same (within errors) for the two systems with the same product $na$ (triangles).
(b) Plotting $\Delta t$ versus $na$, we find $\Delta t\propto(na)^{-2.0(1)}$ (dashed line), matching the prediction $\Delta t\propto(na)^{-2}$.}
\label{fig:3}
\end{figure}

In Fig.~\ref{fig:3}, we study the characteristic time for the WWT particle transport. For a range of $n$ and $a$ (as in Fig.~\ref{fig:2}), we extract the time, $\Delta t(n,a)$, for $\kp$ to drop from $k_0 \approx 1.9\,\um^{-1}$ to $\approx k_0/2$, which is still larger than the largest $\kxi$ here. 

In Fig.~\ref{fig:3}(a), we show that $N_k/N$ curves are the same for $\kp \approx k_0/2$ and different $n$ and $a$ (colored symbols). This means that, as $\kp$ decreases to $k_0/2$, the dynamics of all spectral features have the same dependence on $n$ and $a$, which is thus fully captured by $\Delta t$. Moreover, $\Delta t$ is the same (within errors) for the two systems with the same product $na$, as expected for WWT. In Fig.~\ref{fig:3}(b), we plot $\Delta t$ versus $na$ and find good agreement with the WWT prediction $\Delta t \propto (na)^{-2}$; fitting $\Delta t \propto (na)^{-\alpha}$ (dashed line) gives $\alpha = 2.0(1)$.

\begin{figure}[t!]
\centering
\includegraphics[width=\linewidth]{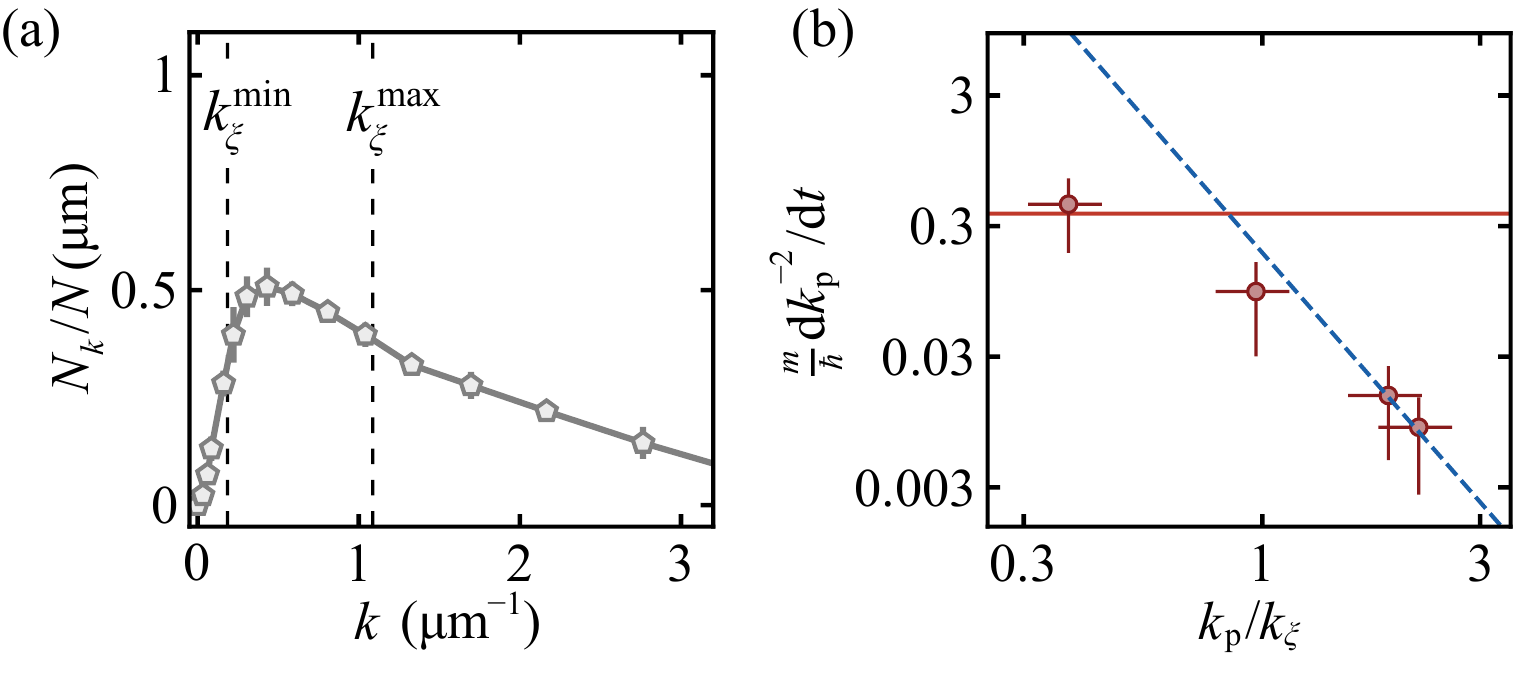}
\caption{
{
Connection to universal coarsening. In WWT, $\upd \kp^{-2}/\upd t$ is $\propto (na)^2 \propto \kxi^4$. However, for $\kp \ll \kxi$, in the coarsening regime, it should be bounded by the universal speed limit, $\approx 0.37\,\hbar/m$ (see text), where $m$ is the atom mass. (a) Here we first prepare a gas with $n\approx 3.7\, \um^{-3}$ and $a=90\,a_0$ in a state with $\kp\approx0.4\,\um^{-1}$, by evolving it  from our initial state with $\kp\approx1.9\,\um^{-1}$, and then change $a$ to different values in the range $(7\textup{--}240)\,a_0$, such that $\kxi$ varies from $\approx 0.19 \,\um^{-1}$ to $\approx 1.1 \,\,\um^{-1}$ (dashed lines).
(b) Extracting $\upd \kp^{-2}/\upd t$ after the change of $a$ and plotting it versus $\kp/\kxi$, we see how it bends away from the WWT scaling (dashed line) towards its speed limit (solid line).}
}
\label{fig:4}
\end{figure}

Finally, in Fig.~\ref{fig:4} we look at how the regime of WWT connects to the universal coarsening observed at $k < \kxi$ once $\kp$ drops well below $\kxi$~\cite{Martirosyan:2025}. In the coarsening regime, which is fully described by the growth of the coherence length, $\kp$ is $\propto 1/\ell$, and ${\rm d}\ell^2/{\rm d}t$ saturates at an $n$- and $a$-independent speed limit, $\approx 3.4\,\hbar/m$, where $m$ is the atom mass; from the data in Ref.~\cite{Martirosyan:2025}, we get $\kp^{-2} \approx 0.11\,\ell^2$, corresponding to ${\rm d}\kp^{-2}/{\rm d}t \approx 0.37\,\hbar/m$. On the other hand, in the WWT regime, ${\rm d}\kp^{-2}/{\rm d}t \propto (na)^2$. We observe both regimes for the same $N_k$ but different $\kp/\kxi$, by first evolving the gas at $90\,a_0$ from our initial state to $\kp \approx 0.4\,\um^{-1}$~[see Fig.~\ref{fig:4}(a)], then switching $a/a_0$ to $7$, $10$, $35$ or $240$, and then extracting ${\rm d}\kp^{-2}/{\rm d}t$~\footnote{We evolve the gas to $\kp \approx 0.4\,\um^{-1}$ always at the same $a=90a_0$ to ensure that we extract ${\rm d}\kp^{-2}/{\rm d}t$ always starting with the same $N_k$, with only $\kp/\kxi$ being different. If we evolve from $k_0$ at different $a=(7\textup{--}240)a_0$, the $N_k$ profiles for the same $\kp \approx 0.4\,\um^{-1}$ and different $a$ are not the same, because for the largest-$a$ system $\kp$ drops well below $\kxi$ already during this evolution~(see also~\cite{Martirosyan:2025}). 
}.
Plotting this ${\rm d}\kp^{-2}/{\rm d}t$ versus $\kp/\kxi$, in Fig.~\ref{fig:4}(b), we show how it bends away from the WWT scaling, $(\kp/\kxi)^{-4}$ (dashed line), towards the coarsening speed limit, $\approx 0.37\,\hbar/m$ (solid line)~\footnote{Note that in Fig.~\ref{fig:3}, $\Delta t$ for the largest $na$ corresponds to $\kp/\kxi$ evolving from $\approx 2.8$ to $\approx 1.4$, and for all the other datasets $\kp/\kxi$ is always $> 2$.}.

In conclusion, we have observed weak wave turbulence as an antecedent to universal coarsening during far-from-equilibrium Bose--Einstein condensation.
For a range of gas densities and interaction strengths, we robustly observe its two hallmark features: the power-law cascade spectrum with exponent $\gamma = 2.4(1)$ and the characteristic particle-transport times $\propto (na)^{-2}$.
In the future, it would be interesting to extend our study to relaxation dynamics at higher per-particle energy, near criticality, and explore the interplay of wave turbulence and near-equilibrium critical fluctuations~\cite{Hohenberg:1977}.

We thank Andrey Karailiev, Christopher J. Ho, Timon A. Hilker, and Vladimir Rosenhaus for useful discussions.
This work was supported by ERC [UniFlat], EPSRC [Grant No.~EP/Y01510X/1], and STFC [Grants No.~ST/T006056/1 and No.~ST/Y004469/1].
Z.~H. acknowledges support from the Royal Society Wolfson Fellowship.

\end{document}